\documentclass[11pt,a4paper]{article} 
\usepackage[english]{babel}
\usepackage{amsmath, amsfonts, amssymb}
\usepackage[]{graphicx}
\usepackage{subfig}
\usepackage[scale=0.7]{geometry}
\usepackage[figurename=Figure, tablename=Table, labelfont=bf]{caption}
\usepackage[colorlinks]{hyperref}
\usepackage{multicol}
\usepackage{xcolor}
\usepackage{colortbl}
\usepackage{hypcap}
\usepackage{array}
\usepackage{siunitx}
\usepackage{authblk}
\usepackage{slashed}
\usepackage{braket}
\usepackage{cite}

\hypersetup{
 pdfborder={0 0 0},
 breaklinks=true,
 linkcolor=blue,
 menucolor=blue,
 urlcolor=blue,
 citecolor=cyan,
 colorlinks=true,
 pdffitwindow=true,
}
\begin{document}

\title{\bf{Chiral constraints on the isoscalar electromagnetic spectral functions of the nucleon from leading order vector meson couplings}}
\author[1,2]{Y. \"Unal}
\affil[1]{\textit{Helmholtz-Institut f\"ur Strahlen- und Kernphysik and Bethe Center for Theoretical Physics, Universit\"at Bonn, D-53115 Bonn, Germany}}
\affil[2]{\textit{Physics Department, \c{C}anakkale  Onsekiz Mart University, 17100 \c{C}anakkale, Turkey}}
\author[1,3,4]{Ulf-G. Mei{\ss}ner}
\affil[3]{\textit{Institute for Advanced Simulation, Institut f\"ur Kernphysik,
and J\"ulich Center for Hadron Physics, Forschungszentrum J\"ulich, D-52425 J\"ulich, Germany}}
\affil[4]{\textit{Tbilisi State University, 0186 Tbilisi, Georgia}}
\date{\empty}

\maketitle

\begin{abstract}
  Using baryon chiral perturbation theory including vector mesons, we analyse various continuum contributions
  to the isoscalar electromagnetic  spectral functions of the nucleon induced by the leading order couplings.
\end{abstract}

\bigskip

\pagebreak

\section{Introduction}

A premier tool to analyse the electromagnetic structure of the nucleon are dispersion relations.
The physics is encoded in the so-called spectral functions, which feature vector meson resonances and
continuum contributions. Much is known about the continuum contribution to the isovector spectral
function of the electric and magnetic nucleon form factors, see e.g. Ref.~\cite{Hoferichter:2016duk}
for the most recent and most precise investigation. The situation for the corresponding continuum
contributions to the isoscalar spectral functions is different, as discussed in detail in
Ref.~\cite{Belushkin:2006qa}. Chiral perturbation theory has been used to analyze the three-pion
continuum~\cite{Bernard:1996cc,Kaiser:2019irl} within the threshold region starting at $t_0 = 9M_\pi^2$.
These two-loop contributions, however, turn out to be rather small, because the
potential enhancement from the anomalous threshold  at $t \simeq 8.9 M_\pi^2$ is efficiently masked by
phase space factors. On the other hand, earlier modeling suggests an important contribution from
the $\pi\rho$ continuum~\cite{Meissner:1997qt} and dispersion relations have been used to
calculate the $K\bar K$ contribution~\cite{Hammer:1999uf,Hammer:1998rz}. In the sector with strange quarks,
there have been suggestions of cancellation between contributios from $K\bar K$, $KK^*$ and $K^*K^*$
loops, although these have very different thresholds~\cite{Barz:1998ih}.

In this paper, we want to re-analyze some of these continuum  as well as some vector
meson contributions to the isoscalar electromagnetic
spectral functions based on chiral perturbation theory with explicit vector mesons. This extends
the earlier tree-level analyses of vector meson contributions to the nucleon electromagnetic (em)
form factors as given e.g. in Refs.~\cite{Kubis:2000zd,Schindler:2005ke}. Here, we work at leading
one-loop order and consider the vector coupling induced contributions. The paper is organized
as follows: In Sec.~\ref{sec:def}, we briefly recall the pertinent definitions for the electromagnetic
form factors and their spectral functions. The $\pi\rho$ contribution is worked out and discussed
in Sec.~\ref{sec:pirho}. The contributions involving $K$ and $K^*$ mesons are given in Sec.~\ref{sec:KK}.
We end with a summary and outlook.  Some technicalities are given in the Appendix.

\section{Isocalar electromagnetic spectral functions}
\label{sec:def}

The electromagnetic form factors of the nucleon are defined via the matrix element of the
electromagnetic current,
\begin{equation}
  \Braket{N(p_{f})|\bar{q}\gamma^\mu Q q|N(p_i)}
  =\;\bar{u}(p_{f})\Big[\gamma^{\mu} F_1(t)+\frac{i}{2m_N}\sigma^{\mu \nu}(p_f-p_i)_\nu F_2(t)\Big]u(p_i)~,
\label{ff}
\end{equation}
with $t=(p_f-p_i)^2$  the invariant momentum transfer squared, $m_N$ the nucleon
mass and $Q$ the quark charge matrix. Throughout, we work in the isospin limit $m_N =(m_p+m_n)/2$, 
with $m_p~(m_n)$ the proton   (neutron) mass.
The functions $F_1(t)$ and $F_2(t)$ are the Dirac and Pauli form factors of the nucleon, respectively.
In the isospin basis, they can be decomposed into
isoscalar and isovector components, following the notation of Ref.~\cite{Mergell:1995bf}
\begin{eqnarray}
F_{i}(t)=F_{i}(t)^S+\tau^3 F_{i}^V(t)~, \quad \quad \quad i=1,2. \nonumber
\end{eqnarray}
$F_1(t)$ and $F_2(t)$ are related to the commonly used  electric and magnetic Sachs form factors
in the following fashion
\begin{equation}
\begin{aligned}
G_E^{S,V}(t)=
&\; F_1^{S,V}(t)+\frac{t}{4 m_N^2} F_2^{S,V}(t)~,   \\
G_M^{S,V}(t)=
&\; F_1^{S,V}(t)+F_2^{S,V}(t)~.
\end{aligned}
\end{equation}
An unsubstracted dispersion relation can be written down for  each form factor defined by
\begin{equation}
F(t)= \frac{1}{\pi} \int_{t_0}^{\infty} \frac{\text{Im} F(t')}{t'-t-i \epsilon}
\end{equation}
where $t_0$ is the threshold energy for hadronic intermediate states. Here, we focus entirely on the
isoscalar spectral function with $t_0=9M_\pi^2$, as the three-pion state is the lowest mass intermediate
state possible. For a more general discussion of the spectral functions, see e.g.
Refs.~\cite{Belushkin:2006qa,Lorenz:2014yda,Leupold:2017ngs}.

\section{The {\boldmath$\pi\rho$} contribution}
\label{sec:pirho}

In this section we focus on the calculation of the imaginary parts of the isoscalar nucleon
electromagnetic form factors generetad from the $\pi\rho$ continuum based on relativistic
two-flavor baryon chiral perturbation theory. At lowest order in the quark mass and momentum expansion,
the relevant interaction Lagrangians are given by \cite{Ecker:1989yg, Ecker:1988te, Gasser:1987rb}
\begin{equation}
\begin{aligned}
\mathcal{L}_{\pi N}=
&\;   \frac{g_A}{2}\bar{\Psi}\gamma^{\mu}\gamma_{5}u_{\mu}\Psi~,   \\
\mathcal{L}_{\rho N}=
&\;   g_{\rho N}\bar{\Psi}\rho_{\mu}\gamma^{\mu}\Psi~,   \\
\mathcal{L}_{\omega \rho\pi}=
&\;   \frac{g_{\omega \rho\pi}}{2}\epsilon^{\mu \nu \alpha \beta} \omega_{\nu}\langle\rho_{\alpha \beta}u_\mu\rangle~,   \\
\mathcal{L}_{\omega \gamma}=
&\;   \frac{f_\omega}{2}e(\partial_{\mu} \omega _{\nu}-\partial_{\nu}\omega_{\mu}) (\partial ^{\mu}  \mathcal{A}^{\nu} - \partial ^{\nu}\mathcal{A}^{\mu})~.
\end{aligned}
\label{L}
\end{equation}
Here, $\Psi$ denotes the nucleon doublet, $\rho_\mu$ the isovector-vector $\rho$-meson, $\omega_\mu$
the isoscalar-vector $\omega$-meson, ${\cal A}^\mu$ the photon field, $\langle \ldots \rangle$ denotes
the trace in flavor space and $\epsilon^{\mu \nu \alpha \beta}$ is the totally antisymmetric Levi-Civita tensor
with $\epsilon^{0123} = 1$. Further, $g_A$ is the nucleon axial-vector coupling, $g_A \simeq 1.27$, and the
various vector meson couplings are taken as $g_{\rho N} = 5.92$, $g_{\omega \rho\pi} =11.6$ \cite{Nakayama:2006ps} and
$f_\omega =0.1$ \cite{Bauer:2012pv}. In the SU$(2)$ sector, due to the universality of the $\rho$-meson coupling
the equality of the $\rho$-meson self-coupling and the coupling to the nucleons $g= g_{\rho N}$, holds \cite{Djukanovic:2004mm}. 
As can be seen from the Lagrangian in Eq.~(\ref{L}), we work in the vector meson dominace approximation, i.e. the
photon couples only via vector mesons to the hadrons. This is e.g. realized in the hidden symmetry approach
of Ref.~\cite{Bando:1987br} for the parameter $a=2$. However, in general, the gauged Wess-Zumino-Witten term
also leads to a direct $\gamma \rho\pi$ coupling e.g. in the massive Yang-Mills approach. It can, however, be
shown that all these different approaches are equivalent, see e.g. the detailed discussion in Ref.~\cite{Meissner:1987ge}.
Our choice is driven by simplicity.
For more thorough discussion  of the Wess-Zumino-Witten term in the presence of
vector mesons and photons, see the reviews~\cite{Bando:1987br,Meissner:1987ge}. Note also that we neglect the small OZI-violating
$\pi$-$\rho$ contribution due to the $\phi$-meson coupling.

\begin{figure}[t!]
\centering
\includegraphics[width=0.8\textwidth]{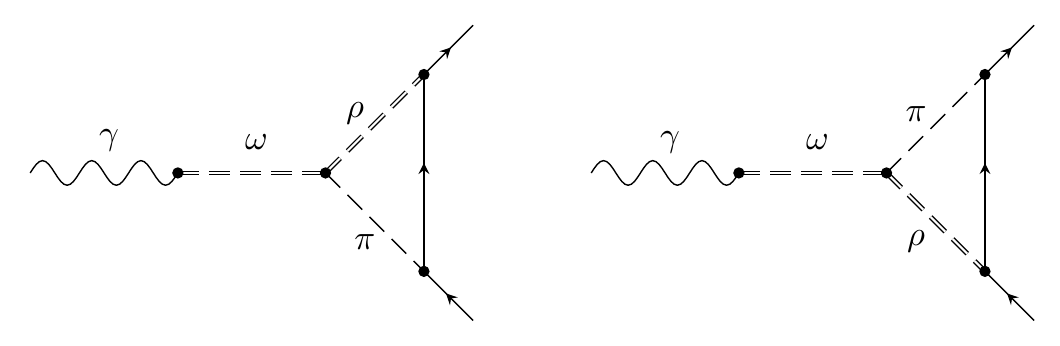}
\caption{Continuum $\pi\rho$ contribution to the imaginary parts of the isoscalar electromagnetic
  form factors of the nucleon through the anomalous coupling to the $\omega$. Dashed and solid lines
  denote pions and nucleons, respectively. The dashed lines represent the $\omega$-  and $\rho$-mesons.}
\label{fig:pirho}
\end{figure}

The following power counting rules for the loop diagrams are used: vertices from $\mathcal{L}^{(n)}$ count
as $\mathcal{O}(q^n)$, so we count the vector meson, nucleon and pion propagators as
$\mathcal{O}(q^0)$, $\mathcal{O}(q^{-1})$ and $\mathcal{O}(q^{-2})$, in order.
Thus the order of the diagrams in Fig.~\ref{fig:pirho} is $\mathcal{O}(q^7)$ at low energies, i.e. for small $t$.

\begin{figure}[t!]
\includegraphics[width=.99\textwidth]{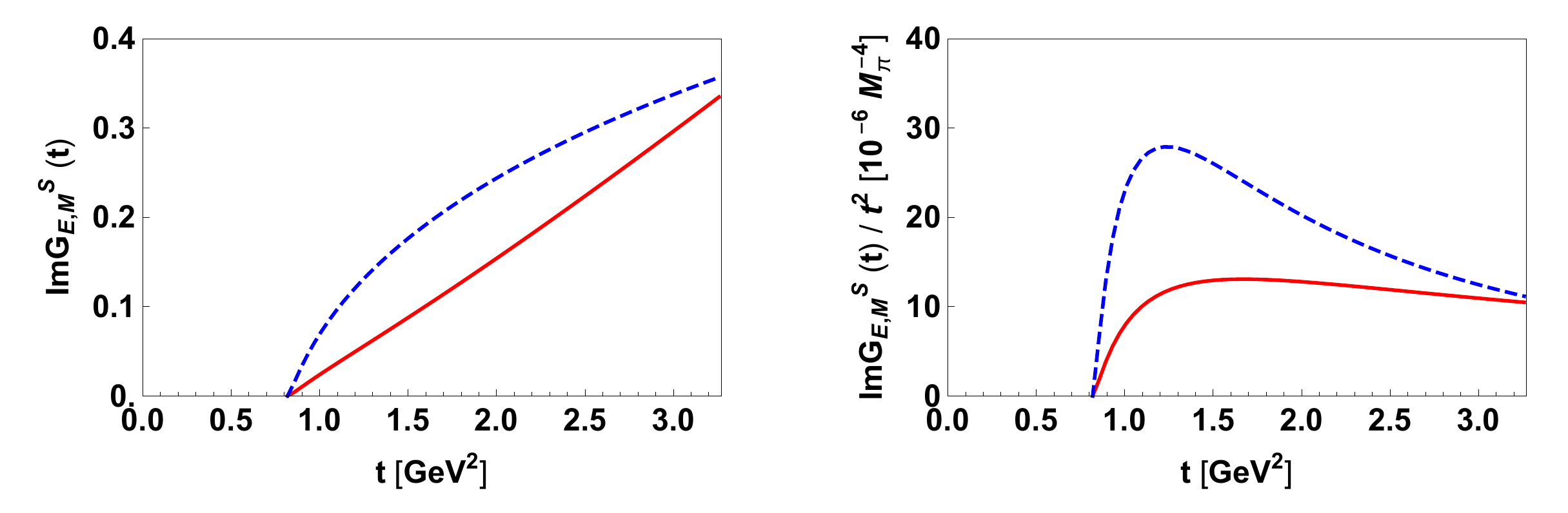}
\centering
\caption{Continuum $\pi\rho$ contribution through the anomalous coupling to the $\omega$
  to the isoscalar spectral functions of the electric and magnetic nucleon form factors
  weighted with and without $1/t^2$. The red
  (solid) and blue (dashed) lines refer to the electric and magnetic form factors,
  $\text{Im}~G_{E}^{S}(t)$ and $\text{Im}~G_{M}^{S}(t)$, respectively.}
  \label{fig:specpirho}
\end{figure}

We work in the center of momentum frame of the nucleon pair with 
$q=\big(-2E_p,~\vec{0}\big)$. The initial and the final momentum of the nucleons are, respectively, 
$p_i=\big(E_p,~\vec{p}\big)$, $p_f=\big(E_p,~\vec{p}\big)$, with $|\vec{p}|= (t/4 - m_N^2)^{1/2}$ and 
$E_p = (m_N^2+|\vec{p}|^2)^{1/2}$.
To calculate the imaginary part of the amplitude for the diagrams which are shown in Fig.~\ref{fig:pirho},
the Cutkosky rules are applied. The imaginary part of the loop diagram corresponds to a cut diagram for
the momentum transfer squared $t \geq(M_\rho+M_\pi)^2$. For this calculation, we perform a reduction
to scalar loop integrals and thus require the basic scalar loop integrals of two- and three- point functions,
called $H$ and $F$, respectively,
\begin{equation}
\begin{aligned}
H=
&\;   i \int \frac{d^4k}{(2\pi)^{4}}\frac{1}{[k^2-{M_{\pi}}^2
       +i\epsilon^+][(k+q)^2-{M_{\rho}}^2+i\epsilon^+]},   \\
F=
&\;   i \int \frac{d^4k}{(2\pi)^{4}}\frac{1}{[k^2-{M_{\pi}}^2
       +i\epsilon^+][(p_i-k)^2-{m_{N}}^2+i\epsilon^+][(k+q)^2
       -{M_{\rho}}^2+i\epsilon^+]}
\end{aligned}
\label{Int}
\end{equation}
with $q^2=t=(p_f-p_i)$ and $\epsilon^+$ stands for $\epsilon \to 0^+$. 
The corresponding imaginary parts can be readily determined
\begin{equation}
\begin{aligned}
\text{Im}~H:\mathcal{H}(t)=
&\;   \frac{M_\pi}{8 \pi \sqrt{t}}\sqrt{\frac{(M_{\rho}^2-M_{\pi}^2-t)^2}{4 M_{\pi}^2 t}-1}~,   \\
\text{Im}~F:\mathcal{F}(t)=
&\;   \frac{1}{16 \pi \sqrt{t}\sqrt{t-4 m_N^2}}\;   
        \log \Bigg[\frac{\frac{t-M_{\pi}^2-M_{\rho}^2}{\sqrt{t-4 m_N^2}}
        -\sqrt{\frac{M_{\pi}^4 + (M_{\rho}^2 - t)^2 - 2 M_{\pi}^2 (M_{\rho}^2 + t)}{t}}}
       {\frac{t-M_{\pi}^2-M_{\rho}^2}{\sqrt{t-4 m_N^2}}
       +\sqrt{\frac{M_{\pi}^4 + (M_{\rho}^2 - t)^2 - 2 M_{\pi}^2 (M_{\rho}^2 + t)}{t}}}\Bigg]~.
\end{aligned}
\label{Imp}
\end{equation}
From these, the expressions for the imaginary part of the isoscalar electric and magnetic
form factor can be given as follows
\begin{eqnarray}
\text{Im}~G_E^{S}(t) &=& \frac{3 f_\omega g_{\omega \rho \pi} g_{\rho N} g_A }{2 \pi^2 F_{\pi}^2}
         \frac{t}{(t-m_{\omega}^2)(t-4 m_N^2)}\nonumber   \\    
         &\times& \Bigg[ 2(t^2 m_N^2+t M_\pi^2M_\rho^2-2 t M_\pi^2m_N^2-2 t M_\rho^2m_N^2 \nonumber  \\
           &+&M_\pi^4m_N^2+M_\rho^4m_N^2-2M_\pi^2M_\rho^2m_N^2) \mathcal{F}(t)
           +2 t (t-M_\pi^2-M_\rho^2)\mathcal{H}(t)\Bigg]~,\\
\text{Im}~G_M^{S}(t) &=&
        \frac{3 f_\omega g_{\omega \rho \pi} g_{\rho N} g_A }{16 \pi^2 F_{\pi}^2 }
        \frac{m_N^2}{(t-m_{\omega}^2)(t-4 m_N^2)^2}\nonumber   \\
        &\times&\Bigg[(2 t m_N^2 (3M_\pi^4+ 3M_\rho^4 -2 M_\pi^2 M_\rho^2 + 8 M_\pi^2 m_N^2
          + 8 M_\rho^2 m_N^2 ) \nonumber\\
         &-& t^2( M_\pi^4 + M_\rho^4 + 8m_N^4 - 4 M_\pi^2 m_N^2  + 12 M_\rho^2 m_N^2 )
         + 2t^3(M_\pi^2 + M_\rho^2 + 3 m_N^2) - t^4 \nonumber \\
         &-& 8m_N^4(M_\pi^4 + M_\rho^4 - 2M_\pi^2 M_\rho^2 ))\mathcal{F}(t)   
        - (t^2+t M_\pi^2 + t M_\rho^2 )(t+8m_N^2) \mathcal{H} (t) \Bigg]~.
\end{eqnarray}
The resulting electric and magnetic spectral functions ${\rm Im}~G^S_{E,M}(t)$ and the weighted spectral functions
 ${\rm Im}~G^S_{E,M}(t)/t^2$ are shown in Fig.~\ref{fig:specpirho}.  The magnetic one shows
 a peak at $t \simeq 1.1\,$GeV$^2$, which is  consistent with the modelling in Ref.~\cite{Meissner:1997qt},
 where the calculated $\pi\rho$ distribution was approximated by a sharp vector meson poles with a mass of
 about 1.1~GeV. For a more detailed comparison, we would also have to include the $\rho$-meson tensor
couplings that appear at next-to-leading order. 

\section{The  contribution from strange loops}
\label{sec:KK}

Next, we consider the effect of loops with $K\bar K$, $K K^*$ and $K^*K^*$ loops. We note, however, that the
latter two only start to have an imaginary part  at about 2~GeV$^2$ and 3~GeV$^2$, which is already
far inside the resonance dominated region. The relevant effective Lagrangians in the SU(3) basis to obtain
the contributions from $K \bar{K}$, $K K^*$ and $K^*K^*$ loops read
\cite{Bando:1987br, Meissner:1987ge, Birse:1996hd, Borasoy:1995ds, Krause:1990xc,Bramon:1992ki}
\begin{equation}
\begin{aligned}
\mathcal{L}_{V \Phi \Phi}=
&\;   -i g \langle V^\mu [\Phi, \partial_\mu \Phi] \rangle,   \\    
\mathcal{L}_{VV\Phi}=
&\;   \frac{1}{\sqrt{2}} G \epsilon^{\mu \nu \alpha \beta}
        \langle \partial_\mu V_\nu \partial_\alpha V_\beta \Phi \rangle,   \\   
\mathcal{L}_{VVV}=
&\;   ig \langle (\partial_\mu V_\nu-\partial_\nu V_\mu) V^\mu V^\nu \rangle, \\    
\mathcal{L}_{BB\Phi}=
&\;   -\frac{F}{2} \langle \bar{B} \gamma^\mu \gamma_5 [u_\mu, B] \rangle
        -\frac{D}{2} \langle \bar{B} \gamma^\mu \gamma_5 \{u_\mu, B\} \rangle ,  \\          
\mathcal{L}_{BBV}=
&\;   G_F (\langle \bar{B} \gamma^\mu [\mathcal{V}_\mu, B] \rangle
       + \langle \bar{B} \gamma ^\mu B \rangle \langle V_\mu \rangle)
       + G_D (\langle \bar{B} \gamma^\mu \{\mathcal{V}_\mu, B\} \rangle
       + \langle \bar{B} \gamma ^\mu B \rangle \langle V_\mu \rangle,   \\
 \mathcal{L}_{\phi \gamma}=  
       &\;   \frac{e}{3\sqrt{2}} F_V (\partial_\mu \phi _\nu-\partial_\nu \phi_\mu)
        (\partial^\mu \mathcal{A}^\nu-\partial^\nu \mathcal{A}^\mu)                
\end{aligned}
\end{equation}
where $\mathcal{V_\mu}$ and $V_\mu$ represent the vector meson octet and the vector meson nonet particles,
respectively. We take the couplings as $G_F=g$ and $G_D=0$ which results from the constraint analysis and
the perturbative renormalizability condition on the vector meson octet and baryon octet interaction
of Ref.~\cite{Unal:2015hea}. The low-energy constants $D$ and $F$ can be determined by fitting to the semi-leptonic
baryon decays at tree level. We use the values $F=0.5$ and $D=0.8$ from Ref.~\cite{Borasoy:1998pe}. Throughout,
we use the pion decay constant $F_\pi$ for the leading order meson decay constant.

In addition to the scalar loop integrals in Eq.~(\ref{Imp}), we also need the following functions corresponding
to the $\bar{K} K$, $K^{*} K^{*}$, $\Lambda \bar{K} K$, $\Sigma \bar{K} K$, $\Lambda K^{*} K^{*}$ and
$\Sigma K^{*} K^{*}$ loops with the equal masses
\begin{equation}
\begin{aligned}
P=
&\;   i \int \frac{d^4k}{(2\pi)^{4}}\frac{1}{[k^2-{M_i}^2
       +i\epsilon^+][(k+q)^2-M_{i}^2+i\epsilon^+]},   \\
J=
&\;   i \int \frac{d^4k}{(2\pi)^{4}}\frac{1}{[k^2-M_{i}^2
       +i\epsilon^+][(k+q)^2
       -M_{i}^2+i\epsilon^+][(p_i-k)^2-m_{j}^2+i\epsilon^+]}.
\end{aligned}
\label{Int2}
\end{equation}
The corresponding imaginary parts are determined as
\begin{equation}
\begin{aligned}
\text{Im}~P:\mathcal{P}_{i}(t)=
&\;   -\frac{1}{16 \pi}\sqrt{1-\frac{4 M_i^2}{t}}~,   \\
\text{Im}~J:\mathcal{J}_{j}(t)=
&\;   \frac{1}{16 \pi \sqrt{t}\sqrt{t-4 m_{j}^2}}\;   
        \log \Bigg[\frac{M_{i}^2 - \frac{t}{2} + \frac{1}{2} \sqrt{-4 m_{j}^2 + t} \sqrt{-4 M_{i}^2 + t}}{M_{i}^2 -  \frac{t}{2} - \frac{1}{2} \sqrt{-4 m_{j}^2 + t} \sqrt{-4 M_{i}^2 + t}}  \Bigg]~
\end{aligned}
\label{Imp2}
\end{equation}
where $M_i$ and $m_j$ denote the masses of the $K, K^{*}$ mesons and of the $\Lambda, \Sigma$ baryons, respectively. 

\begin{figure}[t!]
\centering
\includegraphics[width=1.03\textwidth]{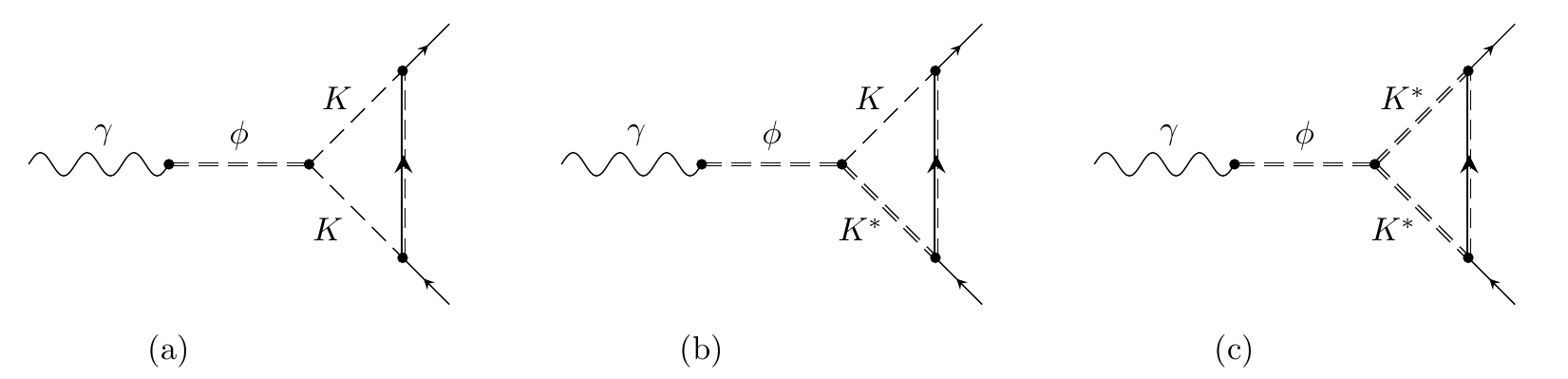}
\caption{Contributions from $K \bar{K}$, $K K^*$ and $K^*K^*$ loops to the imaginary parts of the isoscalar electromagnetic
  form factors of the nucleon through the coupling to the $\phi$. The double wiggly, double dashed and dashed lines
  denote  $\phi$-, $K^{*}$- and  $K$-mesons, respectively. The dashed solid lines represent either 
  $\Lambda$- or $\Sigma$-baryons. There is an additional permutation of the diagrams in group (b).}
\label{fig:KKdiag}
\end{figure} 

We work in the particle basis using the SU(3) symmetric Lagrangian densities given above. An example of such a calculation
can be found in the appendix. The imaginary parts of the Sachs form factors for the $KK$ loop are obtained as
\begin{eqnarray}
\text{Im}~G_E^{S}(t) &=&
     \frac{F_V g}{432 \sqrt{2} F_\pi^2 {\pi}^2 m_N (t-4 m_N^2) (t-M_{\phi}^2)}  \nonumber \\
     &\times& \Bigg[3 (D + 3 F)^2 t (m_\Lambda + m_N)^2 (4 m_{N}^5 -4 M_K^4 m_N- 4 m_{\Lambda}^4 m_N- 2 m_{\Lambda}^2 m_N t \nonumber \\
     &-& 2 m_N^3 t  - m_{\Lambda}^3 (2 t-8 m_N^2) - m_\Lambda (8 m_N^4-6m_N^2 t+ t^2) \nonumber\\
     &+& 2 M_K^2 (4 m_{\Lambda}^2 m_N+ m_N t + m_\Lambda (t-4 m_N^2))) \mathcal{J}_{\Lambda}(t)- 2 t \Big(D^2 (6 m_{\Lambda}^4 m_N \nonumber  \\
     &+& 60 m_N^5 - 108 m_N^3 m_{\Sigma}^2  + 54 m_N m_{\Sigma}^4 + 3 m_{\Lambda}^3 t+ 3 m_\Lambda m_N^2 t -20 m_N^3 t \nonumber   \\ 
     &+&27 m_N^2 m_\Sigma t + 54 m_N m_{\Sigma}^2 t + 27 m_{\Sigma}^3 t + 5 m_N t^2 +6 m_{\Lambda}^2 m_N (t-2 m_N^2)  \nonumber \\
     &-& 2 M_K^2 m_N (3 m_{\Lambda}^2+6 m_\Lambda m_N-10 m_N^2+ 54 m_N m_{\Sigma}+27 m_{\Sigma}+10t)) \nonumber   \\
     &-& 6 D F (-6 m_{\Lambda}^4 m_N + 12 m_N^5 -36 m_N^3 m_{\Sigma}^2 + 18 m_N m_{\Sigma}^4 - 3 m_{\Lambda}^3 t - 3 m_\Lambda m_N^2 t  \nonumber  \\
     &-& 4 m_N^3 t+ 9 m_N^2 m_{\Sigma} t + 18 m_N m_{\Sigma}^2 t + 9 m_{\Sigma}^3 t + m_N t^2 + 6 m_{\Lambda}^2(2m_N^3 -m_N t) \nonumber   \\
     &+& 2 M_K^2 m_N(3 m_{\Lambda}^2 + 6 m_\Lambda m_N + 2 m_N^2 -18 m_N m_\Sigma -9 m_{\Sigma}^2 - 2 t))  \nonumber \\
     &-& 9 F^2 (-6 m_{\Lambda}^4 m_N - 12 m_N^5 + 12 m_N^3 m_{\Sigma}^2 - 6 m_N m_{\Sigma}^4- 3 m_{\Lambda}^3 t - 3 m_{\Lambda} m_N^2 t  \nonumber  \\
     &+& 4 m_N^3 t- 3 m_N^2 m_{\Sigma} t - 6 m_N m_{\Sigma}^2 t - 3 m_{\Sigma}^3 t -  m_N t^2 + 6 m_{\Lambda}^2(2m_N^3-m_N t)  \nonumber \\
     &+& 2 M_K^2 m_N (3 m_{\Lambda}^2 + 6 m_\Lambda m_N - 2 m_N^2 + 6 m_N m_{\Sigma} + 3 m_{\Sigma}^2 + 2 t))\Big)\mathcal{P}_{K}(t))\nonumber   \\
     &-& 27 (D - F)^2 (m_N + m_{\Sigma})^2 (4M_K^4 m_N - 4 m_N^5 + 8 m_N^4 m_{\Sigma} +2 m_N^3 t  \nonumber \\
     &+& 2 m_N m_{\Sigma}^2 (2 m_{\Sigma}^2+t) + m_{\Sigma} t(2m_{\Sigma}^2+t) -2m_N^2(4 m_{\Sigma}^3+3m_{\Sigma} t)\nonumber  \\
     &+& M_K^2(8m_N^2 m_{\Sigma}-2m_{\Sigma}t-2m_N(4m_{\Sigma}^2+t)))\mathcal{J}_{\Sigma}(t)\Bigg]      
\end{eqnarray}
and
\begin{eqnarray}
\text{Im}~G_M^{S}(t) &=&
      \frac{F_V g}{432 \sqrt{2} F_\pi^2 {\pi}^2 (t-4 m_N^2) (t-M_{\phi}^2)}  \nonumber \\
      &\times& \Bigg[6 (D + 3 F)^2 t (m_\Lambda + m_N)^2 (M_K^4 + m_{\Lambda}^4 + m_{N}^4 - 2 M_K^2 (m_{\Lambda}^2 + m_N^2) \nonumber  \\
     &+& m_{\Lambda}^2 ( t-2 m_N^2)) \mathcal{J}_{\Lambda}(t)- t \Big(D^2 (-6 m_{\Lambda}^4 - 12 m_{\Lambda}^3 m_N + 12 m_\Lambda m_N^3 \nonumber  \\
     &+& 60 m_N^4  +  108 m_N^3 m_\Sigma - 108 m_N m_\Sigma^3 - 54 m_\Sigma^4 -3 m_{\Lambda}^2 t \nonumber \\ 
     &-& 6 m_\Lambda m_N t - 70 m_N^2 t - 54 m_N m_\Sigma t - 27 m_{\Sigma}^2 t + 10 t^2  \nonumber \\
     &+& 2 M_K^2 (3 m_{\Lambda}^2 + 6 m_\Lambda m_N + 110 m_N^2 + 54 m_N m_\Sigma + 27 m_{\Sigma}^2 - 20 t))  \nonumber \\
     &+& 6 D F (-6 m_{\Lambda}^4 - 12 m_{\Lambda}^3 m_N + 12 m_{\Lambda} m_N^3 - 12 m_N^4 - 36 m_N^3 m_\Sigma   \nonumber \\
     &+& 36 m_N m_{\Sigma}^3 + 18 m_{\Sigma}^4 - 3 m_{\Lambda}^2 t - 6 m_\Lambda m_N t + 14 m_N^2 t + 18 m_N m_{\Sigma} t  \nonumber  \\
     &+& 9 m_{\Sigma}^2 t - 2 t^2 + 2 M_K^2 (3 m_{\Lambda}^2 + 6 m_\Lambda m_N - 22 m_N^2 - 18 m_N m_{\Sigma} - 9 m_{\Sigma}^2 + 4 t)) \nonumber  \\
     &+& 9 F^2 (-6 m_{\Lambda}^4 - 12 m_{\Lambda}^3 m_N + 12 m_{\Lambda} m_N^3 + 12 m_N^4 + 12 m_N^3 m_\Sigma - 12 m_N m_{\Sigma}^3  \nonumber  \\
     &-& 6 m_{\Sigma}^4 - 3 m_{\Lambda}^2 t - 6 m_\Lambda m_N t - 14 m_N^2 t - 6 m_N m_\Sigma t - 3 m_{\Sigma}^2 t + 2 t^2 \nonumber  \\
     &+& 2 M_K^2 (3 m_{\Lambda}^2 + 6 m_\Lambda m_N + 22 m_N^2 + 6 m_N m_\Sigma + 3 m_{\Sigma}^2 - 4 t)) \Big)\mathcal{P}_{K}(t)) \nonumber  \\
     &+& 54 (D - F)^2 (m_N + m_{\Sigma})^2 (M_K^4 + m_N^4 - 2 m_N^2 m_{\Sigma}^2 + m_{\Sigma}^4\nonumber  \\
     &-& 2 M_K^2 (m_N^2 + m_{\Sigma}^2) + m_{\Sigma}^2 t)\mathcal{J}_{\Sigma}(t)\Bigg].  
\end{eqnarray}
\begin{figure}[t]
\includegraphics[width=.99\textwidth]{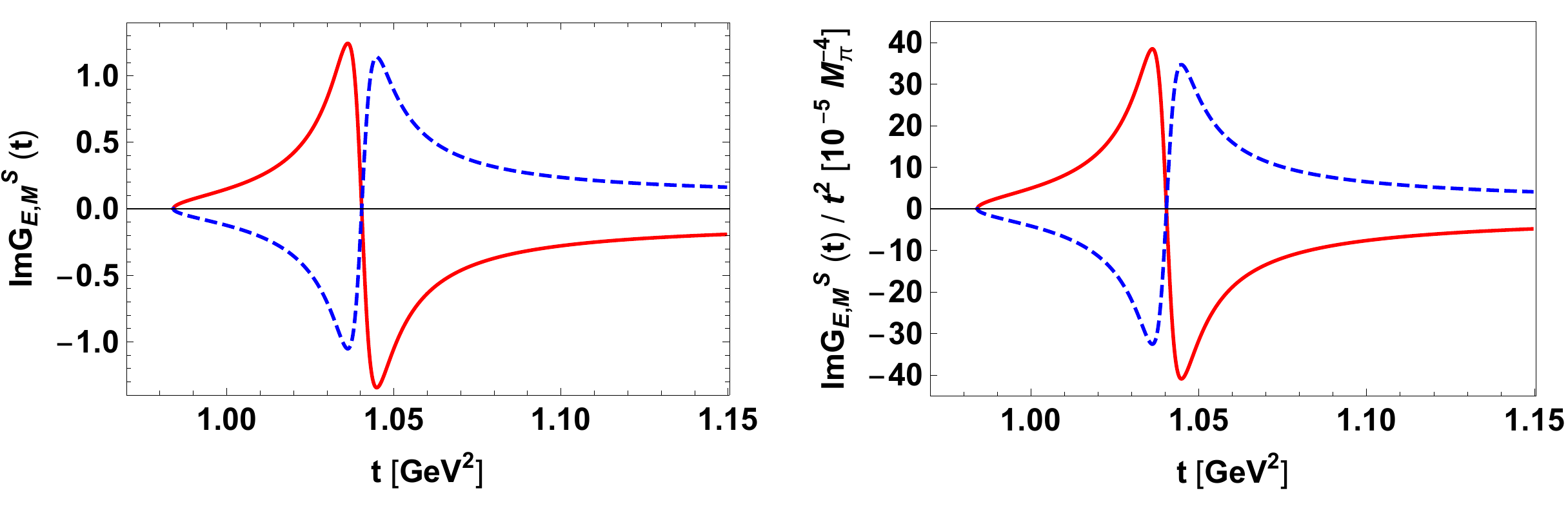}
\centering
\caption{$K\bar K$ contribution through the anomalous coupling to the $\phi$
  to the isoscalar spectral functions 
  weighted with and without $1/t^2$. The red
  (solid) and blue (dashed) lines refer to the electric and magnetic form factors,
  $\text{Im}~G_{E}^{S}(t)$ and $\text{Im}~G_{M}^{S}(t)$, respectively.}
\label{fig:KK}
\end{figure} 
\noindent
The definitions of $\mathcal{J}_{\Lambda}(t)$, $\mathcal{J}_{\Sigma}(t)$ and $\mathcal{P}_{K}(t)$ can be found in Eq.~(\ref{Imp2}).
In Fig.~\ref{fig:KK}, we show the $K\bar K$ loop contribution including the the one from the $\phi$-meson. To account for its
finite width, we have substituted $M_\phi$ by $M_\phi - i\Gamma_\phi/2$, where $\Gamma_\phi \simeq 4\,$MeV is the width of the
$\phi$. We find similar spectral functions for the electric and magnetic form factors except for the sign. Similar to Ref.~\cite{Hammer:1999uf},
there is no enhancement on the left wing of the $\phi$ which sits directly at the $K\bar K$ threshold. Our results with vector mesons
at one loop can, however, not directly be compared to dispersion-theoretical result of Ref.~\cite{Hammer:1998rz} as their calculation
does not include all of the $\phi$-meson strength and ours would also require the inclusion of the NLO tensor coupling. Still, it is
comforting to see that in the region of the $\phi$ there is little continuum strength.
As noted earlier, the spectral functions generated by $K^*K$ and $K^*K^*$ start at much larger $t$
and turn out to be rather small, see Fig.~\ref{fig:KsK} and Fig.~\ref{fig:KsKs}, respectively.

\begin{figure}[t]
\includegraphics[width=.95\textwidth]{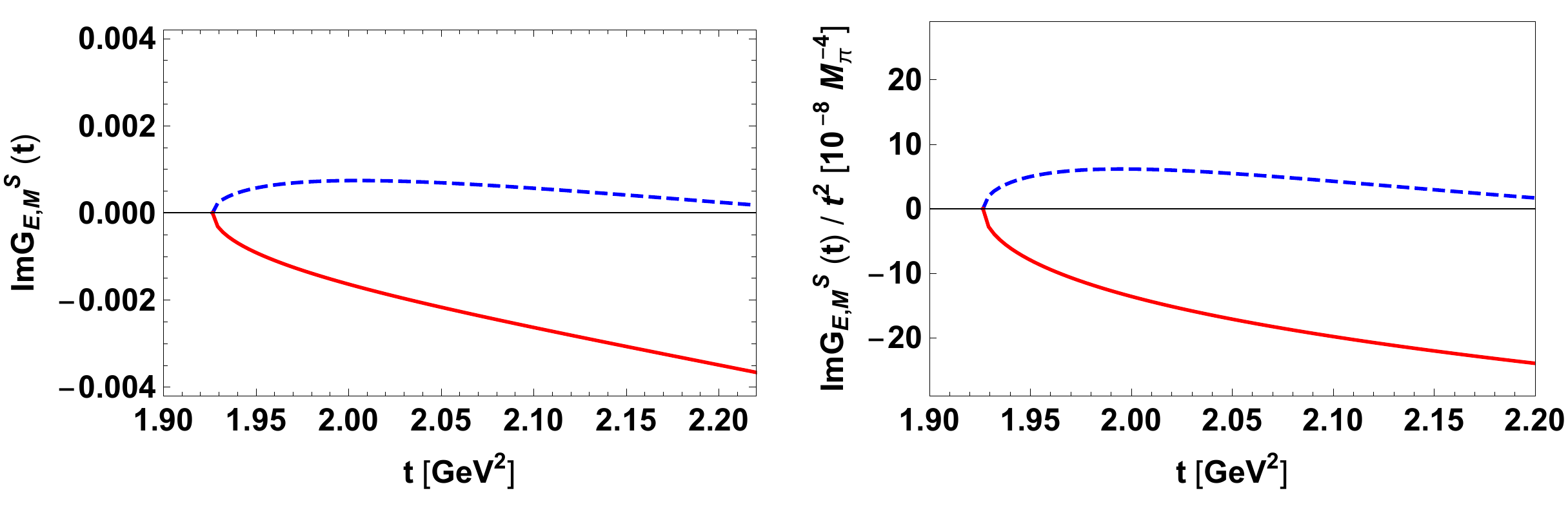}
\centering
\caption{$K^{*}K$ contribution through the anomalous coupling to the $\phi$
  to the isoscalar spectral functions 
  weighted with and without $1/t^2$. For notations, see Fig.~\ref{fig:KK}.}
\label{fig:KsK}
\end{figure} 

\vspace{-3mm}

\begin{figure}[t]
\includegraphics[width=.95\textwidth]{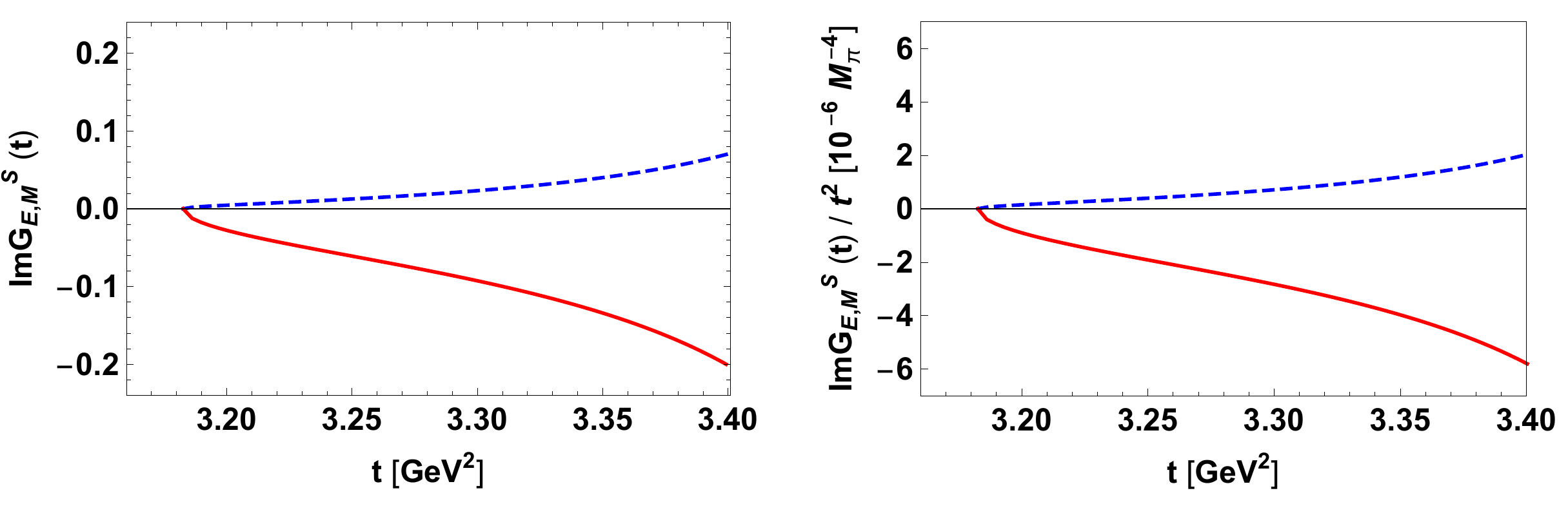}
\centering
\caption{$K^{*}K^{*}$ contribution through the anomalous coupling to the $\phi$
  to the isoscalar spectral functions 
  weighted with and without $1/t^2$. For notations, see Fig.~\ref{fig:KK}.} 
\label{fig:KsKs}
\end{figure} 

\section{Summary}

In this paper, we have used baryon chiral perturbation theory including vector mesons to derive new chiral constraints on the
isoscalar electromagnetic spectral functions based on the leading order vector coupling. As noted earlier, the $\pi$-$\rho$
loop contribution is the most substantial
non-resonant effect and needs to be included. The non-resonant $K\bar K$ contribution that sits under the $\phi$-meson is less
pronounced, which is consistent with earlier findings, used e.g. in Ref.~\cite{Belushkin:2006qa}. The inclusion of the
tensor coupling effects should be considered next.

\section*{Acknowledgments}
We are grateful to Jambul Gegelia for helpful discussions. 
We acknowledge partial financial support from the Deutsche Forschungsgemeinschaft (SFB/TRR~110, ``Symmetries
and the Emergence of Structure in QCD''), by the Chinese 
Academy of Sciences (CAS) President's International Fellowship Initiative (PIFI) (grant no. 2018DM0034) 
and by VolkswagenStiftung (grant no. 93562).

\section*{Appendix: A  sample SU(3) calculation}
\label{app:1}

\begin{figure}[htb!]
\centering
\includegraphics[width=0.5\textwidth]{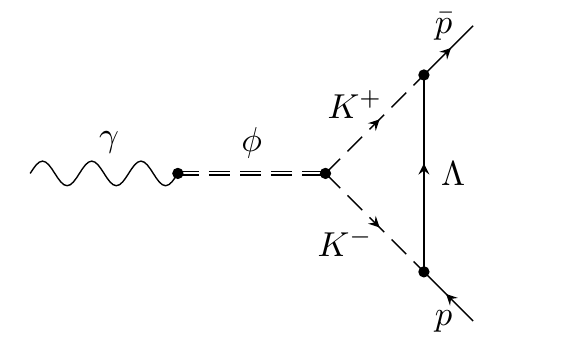}
\caption{A Feynman diagram for the $\bar{K}K\Lambda$ loop contribution.}
\label{fig:exp}
\end{figure} 

As an example for the pertinent SU(3) calculations, we consider the $K\bar K\Lambda$ contribution depicted in Fig.~\ref{fig:exp}.
The Feynman rules for the $\phi \gamma$, $\phi K^{+} K^{-}$, $\Lambda K^{-} p$ and $\Lambda K^{+} \bar{p}$ vertices appearing in this
diagram are obtained as follows
\begin{eqnarray*}
\phi \gamma&:&  \frac{2 i e F_V}{3 \sqrt{2}}  (g^{\mu \nu} q^2-q^\mu q^\nu)   \\ 
\phi K^{+} K^{-}&:&  2(\partial^\mu K^{+}) K^{-} \phi_{\mu}-2(\partial^\mu K^{-}) K^{+} \phi_{\mu}  \\ 
&&       \qquad                i \mathcal{L} \rightarrow  2 ig(k_{1}^\mu-k_{2}^\mu)  \\ 
p K^{-} \bar{\Lambda} &:& -\frac{D}{2\sqrt{3}} \bar{\Lambda}(\partial^\mu K^{-})p-\frac{F \sqrt{3}}{2} \bar{\Lambda}(\partial^\mu K^{-})p \\
&&       \qquad                i \mathcal{L} \rightarrow  \frac{(D+3F)}{2\sqrt{3}F_\pi}\gamma_{5}\slashed{k_2}  \\ 
\Lambda K^{+} \bar{p} &:&  -\frac{D}{2\sqrt{3}} \bar{p}(\partial^\mu K^{+})\Lambda-\frac{F \sqrt{3}}{2} \bar{p}(\partial^\mu K^{+})\Lambda \\
&&       \qquad                i \mathcal{L} \rightarrow  \frac{(D+3F)}{2\sqrt{3}F_\pi}\gamma_{5}\slashed{k_1}  \\ 
\end{eqnarray*}
Here, $k_1$ and $k_2$ denote the momenta for the $K^{+}$ and the $K^{-}$, in order. The $\phi$ vector meson has momentum $q$ and
Lorentz index $\mu$. The amplitude for the diagram in Fig.~\ref{fig:exp} reads
\begin{eqnarray*}
\Gamma^\mu(p_i, p_f, q) &= &  \frac{e g F_V (D+3F)^2}{24 \sqrt{2}F_\pi^2 \pi^4}   \\
&\times& \mu^{4-n}  \int d^nk
\frac{(q^\eta q^\mu-q^2 g^{\eta \mu}) (2 k^\delta-q^\delta) (g_{\delta \eta} M_{\phi}^2-q_\delta q_\eta) \gamma^5(\slashed{q}-\slashed{k})
  (\slashed{p_i}+ \slashed{k}+m_\Lambda)\gamma^5\slashed{k}}{(q^2-M_\phi^2) (k^2-M_K^2) ((k+p_i)^2-m_{\Lambda}^2) ((q-k)^2-M_K^2)},
\end{eqnarray*}
with $n$ the number of dimensions and $\mu$ the scale of dimensional regularization.

\end{document}